\documentclass[journal]{IEEEtran}
\usepackage[cmex10]{amsmath}
\usepackage{amssymb}
\usepackage{amsfonts}
\usepackage{graphicx}
\usepackage{tabularx}
\usepackage{booktabs} 
\usepackage{ifpdf}
\usepackage{epstopdf}
\usepackage{cite}
\usepackage{algorithmic}
\usepackage{color}
\usepackage{array}
\usepackage{multicol}
\usepackage[caption=false,font=footnotesize]{subfig}
\usepackage[table]{xcolor}

\ifCLASSINFOpdf
\else
\fi

\begin{document}
%
\title{Joint Space–Time Coding on RIS for Simultaneous Direct Modulation and Beamforming}

\author{
{
Shiyuan Li, Baojiang Yan, Yihan Xie, Yixin Tong,
~Chong He,~\IEEEmembership{Member,~IEEE}
~Xudong Bai,~\IEEEmembership{Member,~IEEE}
~Qingqing Wu, ~\IEEEmembership{Senior Member,~IEEE}
~Wen Chen, ~\IEEEmembership{Senior Member,~IEEE}

}
\thanks{
Shiyuan Li, Baojiang Yan, Yihan Xie, Qingqing Wu, Wen Chen, Chong He are with the Department of Electronic Engineering, Shanghai Jiao Tong University, 800 Dongchuan RD. Minhang District, Shanghai, China (e-mail: hechong@sjtu.edu.cn)

Xudong Bai is with the School of Microelectronics, Northwestern Polytech
nical University, Xi’an 710060, China (e-mail: baixudong@nwpu.edu.cn)

Yixin Tong is with the College of Communications Engineering, Army 
Engineering University of PLA, Nanjing 210007, China.
}}

\markboth{Journal of \LaTeX\ Class Files,~Vol.~11, No.~4, December~2025}%
{Shell \MakeLowercase{\textit{et al.}}: Bare Demo of IEEEtran.cls for Journals}

\maketitle

\begin{abstract}
Reconfigurable Intelligent Surface (RIS)-based direct modulation communication systems have garnered significant attention due to their low cost, low power consumption, and baseband-less characteristics. However, these systems face challenges such as the random time-varying coding state of the RIS and the difficulty in implementing beamforming in direct modulation. In this paper, we propose a simple and effective joint space-time coding approach for RIS that enables simultaneous realization of both direct modulation communication and beamforming. By modeling the transmitted signals of the RIS using space-time coding, we show that the time coding determines the direct modulation functionality, while the space coding governs the beamforming. Consequently, we introduce a joint time-space coding technique by performing exclusive-or (XOR) operations on the time and space coding sequences, enabling both functionalities to be achieved concurrently. Numerical simulations demonstrate the effectiveness of the proposed method. Furthermore, we design and fabricate a transmissive 1-bit phase reconfigurable RIS operating in the 3.4-3.79 GHz frequency band for the implementation of a direct modulation communication system. Experimental results reveal that the bit error rate (BER) is significantly reduced when joint space-time coding is used, compared to using time coding alone. Additionally, the root-mean-square error vector magnitude (rmsEVM) of the constellation diagram is reduced by 55\%. This technique is promising for applications in the Internet of Things (IoT), contributing to the development of intelligent networks for electronic devices.
\end{abstract}


%
\IEEEpeerreviewmaketitle

\section{Introduction}

\IEEEPARstart{D}{irect} modulation based on RIS is considered promising owing to its simple hardware architecture and low power consumption \cite{tang2020wireless,M2024wireless}. 
In this approach, the baseband information is directly imposed onto the radio frequency (RF) signal by controlling the phase or amplitude of the electromagnetic (EM) waves through the metasurface \cite{li2024transmissive,10711860}. 
As a result, by appropriately configuring the position and radiation mode of the metasurface, applications such as the Internet of Things (IoT) \cite{gholami2022direct,Van2024Enhancing} and backscatter communications \cite{zhao2020metasurface,xu2022intelligent,Zhang2024Design} can be effectively enabled among different devices and users \cite{Fan2023minia,Li2024Meta}.

RIS originates from the concept of digitally coded metasurfaces \cite{cui2014coding}. 
Digital coding metasurfaces have demonstrated flexible control over multiple dimensions of radio frequency (RF) electromagnetic waves, including phase \cite{10739947}, amplitude\cite{han2024complex}, polarization\cite{liu2024arbitrarily}, and frequency\cite{8115203}. This capability has motivated extensive research on their integration into novel wireless communication transmitters. For example, the phase reconfigurable metasurface can be utilized to replace the mixer and phase-shifter in the traditional system hardware for direct phase-shift keying (PSK) modulation \cite{tang2019programmable, cui2019direct}. Similarly, direct amplitude-shift keying (ASK) modulation can be realized by the metasurface with EM wave amplitude control \cite{Zheng2021Direct, zheng2023spatial, jwair2023intelligent}. At a more advanced level, Qi \emph{et al.} \cite{xiong2024multi} employed an amplitude–phase reconfigurable space-coding metasurface to realize dual-channel independent direct modulation communication. This metasurface provides 3-bit phase modulation and 1-bit amplitude modulation capabilities. Although it offers higher phase resolution, its operational bandwidth is extremely narrow, effectively approaching a single frequency point. In general, these studies primarily map baseband information directly onto the phase or amplitude response of the RIS. In other words, this class of direct modulation communication methods involves modulating the information bitstream directly onto the carrier, which restricts the achievable modulation order to the precision of phase or amplitude control supported by the RIS. For instance, a 1-bit phase-reconfigurable RIS can only support binary PSK (BPSK) modulation. To realize higher-order schemes such as 16QAM, a 2-bit phase-reconfigurable and 2-bit amplitude-reconfigurable RIS would be required, which poses significant challenges for practical RIS design.

\begin{table*}[t]
	\centering
	\renewcommand{\arraystretch}{1.3} 
	\caption{Comparison of representative RIS-based direct modulation schemes and the proposed method}
	\label{tab:RIS_modulation}
	\begin{tabular}{lcccc}
		\hline
		\textbf{Reference} & \textbf{Information carrier} & \textbf{Beamforming} & \textbf{Modulation scheme} & \textbf{RIS} \\
		\hline
		\cite{tang2019programmable} & Carrier frequency & $\times$ & 8PSK & 3-bit phase reconfigurable \\
		\cite{xiong2024multi}       & Carrier frequency & $\checkmark$ & OOK & Independent 3-bit phase and 1-bit amplitude reconfigurable \\
		\cite{dai2019realization}   & $+1$st harmonic   & $\times$ & QPSK, 8PSK, 16QAM & 1-bit phase reconfigurable \\
		\cite{chen2022accurate}     & $+1$st harmonic   & $\times$ & 256QAM & 1-bit phase reconfigurable \\
		\cite{zhang2021wireless}    & $\pm 1$st harmonics & $\times$ & OOK & 2-bit phase reconfigurable \\
		\cite{10785548}             & $\pm 1$st harmonics & $\times$ & FSK & 1-bit phase reconfigurable \\
		\rowcolor{gray!15} This work                   & $+1$st harmonic   & $\checkmark$ & 16QAM & 1-bit phase reconfigurable \\
		\hline
	\end{tabular}
\end{table*}

Fortunately, periodic phase modulation applied to a RIS with only 1-bit phase reconfigurability enables accurate amplitude–phase control of the generated harmonics by adjusting the duty cycle and initial timing of the modulation sequence \cite{zhang2018spacetime,dai2018independent,fang2022design, 11105439}. 
In other words, by modulating the 1-bit phase in time according to specific timing patterns, equivalent multi-bit phase and amplitude control can be achieved at the harmonics. This provides the theoretical foundation for realizing higher-order and multi-dimensional modulation schemes in the harmonic domain, rather than being limited to the carrier frequency. In \cite{dai2019realization}, multi-modulatioin schemes, such as QPSK, 8-PSK and 16QAM , were realized by time-domain digital metasurface with 1-bit reconfigurable phase. By controlling the time-domain coding sequence, the information to be transmitted is mapped to the duty ratio and relative time delay in each symbol period \cite{ni2022reconfigurable}. With accurate manipulation by such time-domain digital coding metasurface, high-order quadrature amplitude modulation (QAM) such as 256QAM \cite{chen2022accurate} can be reached. However, the scheme requires high sampling rate and synchronization of the receiving part to ensure correct time-coding. The aforementioned work utilises solely the $+1$st harmonic for information transmission. In \cite{zhang2021wireless}, a 2-bit phase-reconfigurable RIS was fabricated, where its generated $+1$st and $-1$st harmonics were independently modulated to convey distinct information, thereby achieving dual-channel transmission via harmonics. In \cite{10785548}, simultaneous FSK modulation was achieved on both harmonics by controlling the amplitudes of the $+1$st and $-1$st harmonics generated by a directly radiating 1-bit phase-reconfigurable RIS. Based on the time-coding of metasurface, multiple input multiple output (MIMO) \cite{chen2021design,Zhu2023Modulation} and multistream \cite{li2023scalable} are studied. 

However, these studies on direct modulation communication via harmonics overlook the beamforming gain and flexible beam control capabilities of the array. For periodic time coded metasurfaces, the generated harmonics are stable, and beam scanning at harmonics can be achieved by adjusting the modulation time delay between adjacent elements \cite{tian2022programmable}. In contrast, in direct modulation communication systems based on harmonics, the modulation timing is dictated by the information bitstream, which is typically random rather than periodic. As a result, stable $+1$st harmonic spectral lines are absent \cite{yan2024metasurface}, and conventional harmonic beam-scanning analysis methods no longer apply.

It is noteworthy that for RIS, whether employing periodic or non-periodic time-domain modulation, the equivalent amplitude and phase of the fundamental component remain constant, implying that the spectrum of the fundamental (carrier) component is stable. Moreover, since the symbol rate is much smaller than the carrier frequency, the power radiation pattern corresponding to the carrier frequency can be used to approximate that at the harmonic frequencies. Beamforming at the carrier frequency corresponds to a one-time encoding of the phase of each RIS element, i.e., space coding. To embed the information bitstream onto the harmonics generated by the carrier through time-domain modulation, additional time-domain modulation of the RIS is required. 

Motivated by this observation, this paper proposes a space–time joint coding strategy that simultaneously enables harmonic-based direct modulation communication and beam scanning, thereby enhancing the performance of RIS-based direct modulation communication. Specifically, inspired by the addition theorem of digital coding \cite{wu2018addition}, we investigate the coding sequence of metasurface from time-domain and space-domain, respectively. Then, the joint coding matrix is derived from the digital exclusive OR (XOR) logic operation, which decouples the dependence of the existing space coding sequence and time coding sequence. Consequently, the state switching from time-coding and specific array patterns are obtained concurrently. The comparison of representative RIS-based direct modulation schemes and the proposed method is shown in Table~\ref{tab:RIS_modulation}. The main contributions of this paper are summarized as follows,




(1) To the best of our knowledge, the joint space-time coding design method for enhancing RIS-based direct modulation communications is proposed for the first time in this paper. The innovative space-time joint coding is produced by performing XOR operation on the time coding and the space coding. Based on the novel space-time joint coding, beamforming is realized without affecting the direct modulation communication function, which can effectively improve the SNR at the receiver and reduce the BER.

(2) A direct modulation communication system consisting of a 1-bit phase reconfigurable transmissive RIS is constructed in this paper. The simultaneous realization of beamforming and direct modulation communication using a space-time joint coding RIS is experimentally verified, which increases the SNR at the receiver and reduces the BER.

The rest of this article can be outlined as follows. Section II introduces the principle and theory of the joint space-time-coding calculation and beam steering. Section III provides numeric simulations of the radiation patterns and BER performance. In Section IV, experiments and results are shown to verify the feasibility of communication improvement. Finally, Section V concludes this article and gives potential future improvements.

\section{Theory and System Design}

\begin{figure*}[!t]
	\centering
	\includegraphics[width=10cm]{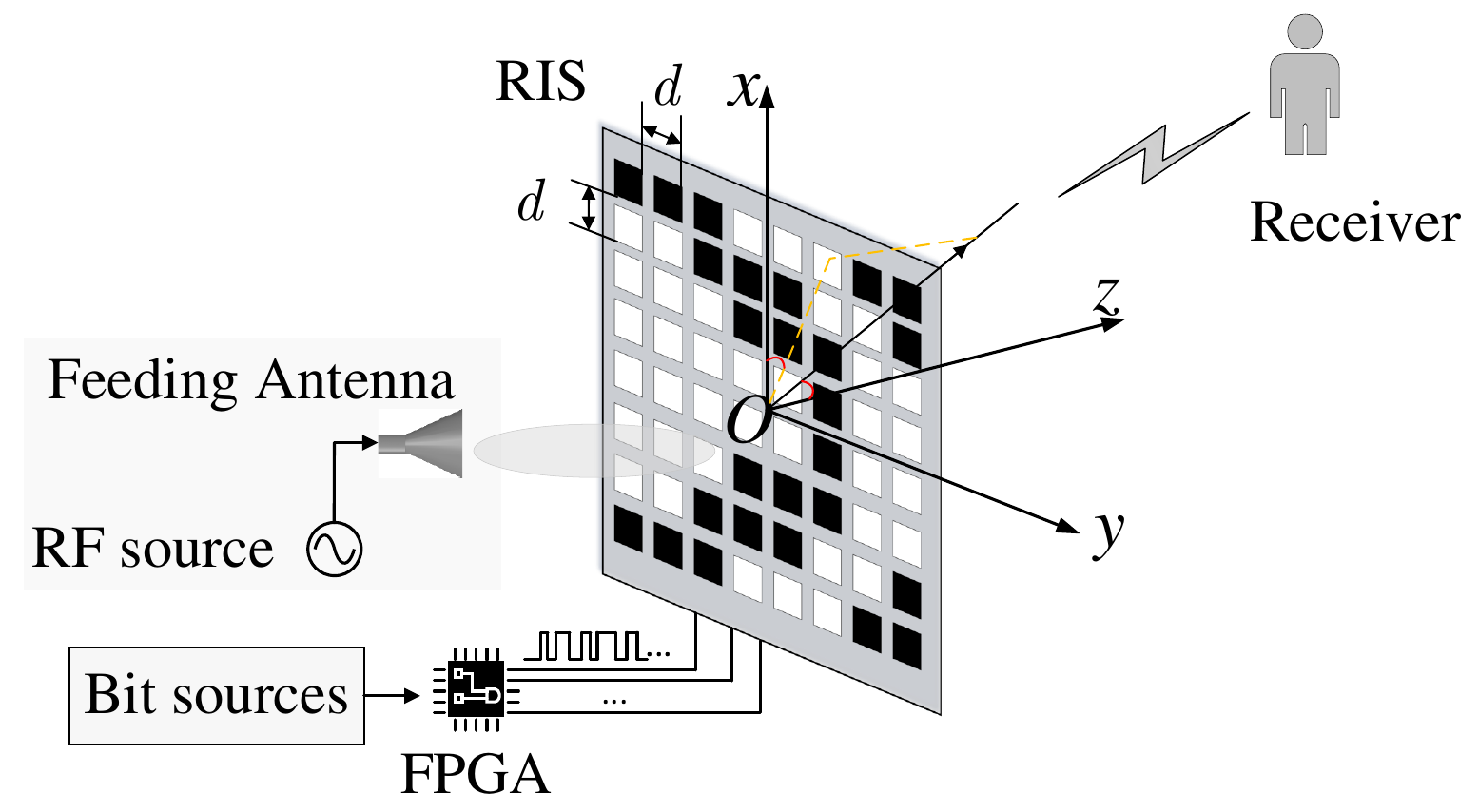}
	\caption{Schematic diagram of enhanced direct modulation communication through joint space-time modulation.}
	\label{Fig1}
\end{figure*}

\begin{figure}[!t]
	\centering
	\includegraphics[width=5cm]{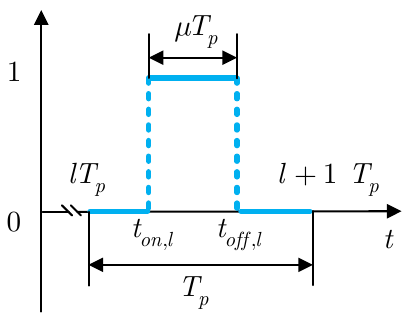}
	\caption{Time sequence $g_l(t)$ of the $l$th modulation period.}
	\label{Fig2}
\end{figure}

This section focuses on the theory and system design of space-time coding for simultaneous beamsteering and direct modulation communication. 
\subsection{Direct Modulation Communication Model}
The direct modulation communication system based on space-time coding RIS is shown in Fig.~\ref{Fig1}, and its transmitter side mainly includes RIS, feeding antenna and field programmable gate array (FPGA). The RIS consists of $M \times N$ units with unit spacing $d$. Take the plane in which the RIS is located as the $xoy$ plane and set up a right-handed Cartesian coordinate system $O-xyz$. The coordinate of the feeding antenna is $(x_c, y_c, z_c)$. The receiver is in the far-field region of the RIS and the orientation with respect to the RIS can be represented by the direction vector $\vec{I}$
\begin{equation}
	\label{dvc}
	\vec{I}=\left( \sin \theta \cos \varphi ,\sin \theta \sin \varphi ,\cos \theta  \right).
\end{equation}

The distance between the feeding antenna and the $(m,n)$th unit can be expressed as
\begin{equation}
	\resizebox{0.89\hsize}{!}{$	
	\label{dismn}
	{{r}_{m,n}}=\sqrt{{{\left[ \left( n-\frac{N+1}{2} \right)d-{{x}_{c}} \right]}^{2}}+{{\left[ \left( m-\frac{M+1}{2} \right)d-{{y}_{c}} \right]}^{2}}+z_{c}^{2}}.
	$}
\end{equation}

The RIS is assumed to be 1-bit phase reconfigurable, with each cell having two states, `ON' and `OFF'. When the unit switches between these two states, the phase of the transmitted electromagnetic wave is shifted by $180^\circ$, while the same transmitted amplitude is maintained. The state of each unit can be controlled by the FPGA output signal $G\left( t \right)$, which can be represented as
\begin{equation}	
		\label{gt}
		G(t)=\sum\limits_{l=-\infty }^{\infty }{{{g}_{l}}(t-}l{{T}_{p}}),
\end{equation}
where $G\left( t\right)$ is the time coding sequences and ${{T}_{p}}$ is the switching period, then the corresponding switching frequency is $F_p = 1/T_p$. The ${{g}_{l}}(t)$ is the output signal of the FPGA in the $l$th period, as shown in the Fig.~\ref{Fig2}, which can be denoted as
\begin{equation}
		\label{glt}
		{{g}_{l}}(t)=\left\{ \begin{aligned}
			& 1,{{t}_{on,l}}<t<\mu {{T}_{p}}+{{t}_{on,l}} \\ 
			& 0,0\le t<{{t}_{\text{on,}l}}\cup \mu T_p+{{t}_{on,l}}<t<{{T}_{p}} \\ 
		\end{aligned} \right.
\end{equation}

The unit of RIS is in `ON' state when ${{g}_{l}}(t)=1$. When ${{g}_{l}}(t)=0$, the unit is in the `OFF' state. Ideally, for a 1-bit phase-reconfigurable RIS, the phase response in the `ON' state differs by $180^\circ$ from that in the `OFF' state. $\mu$ denotes the ratio of the time duration of the 'ON' state to the switching period $T_p$. ${{t}_{on,l}}$ denotes the starting instant in the $l$th period when the unit is in the 'ON' state. Under the control of the modulation timing $g_l(t)$, the phase response of the $(m,n)$th unit of the RIS can be expressed as
\begin{equation}
	\label{hmnlt}
	h_{m,n,l}(t)=
	\begin{cases}
		e^{j\pi}, & t_{on,l} < t < \mu T_{p} + t_{on,l}, \\
		e^{j0}, & 0 \le t < t_{on,l} \;\; \text{or} \;\; \mu T_p + t_{on,l} < t < T_{p}.
	\end{cases}
\end{equation}
Then, applying the discrete Fourier transform to (\ref{hmnlt}) yields,
\begin{equation}
	\label{HK}
	H[kF_p] =
	\begin{cases}
		1 - 2\mu, & k = 0, \\[8pt]
		-\dfrac{2}{\pi k} \, \sin(\pi k \mu) \, e^{-j\left(2\pi k F_p t_{on,l} + \pi k \mu\right)}, & k \neq 0.
	\end{cases}
\end{equation}
The amplitude and phase corresponding to the $k$th harmonic can be expressed respectively as
\begin{align}
	|H[kF_p]| &= \dfrac{2}{\pi |k|}\,\big|\sin(\pi k \mu)\big|, \quad k \neq 0, \\[6pt]
	\angle H[kF_p] &= -\left(2\pi k F_p t_{on,l} + \pi k \mu\right) + 
	\begin{cases}
		0, & \sin(\pi k \mu) \geq 0, \\
		\pi, & \sin(\pi k \mu) < 0.
	\end{cases}
\end{align}
where $|H[kF_p]|$ and $\angle H[kF_p]$ denote the amplitude and phase of the complex number $H[kF_p]$, respectively. This indicates that by adjusting the duty cycle $\mu$ and the initial timing at which the RIS enters the `ON' state, one can control both the amplitude and phase of the generated $k$th harmonic component. Therefore, the amplitude and phase of the generated harmonic components can be exploited for symbol mapping. That is to say, through the time encoding of the RIS, harmonic-based direct modulation communication can be achieved. \cite{cui2019direct,ni2022reconfigurable}. 

As the RIS forms an array, employing beamforming to reduce communication bit error rates is a natural approach. However, because the 0/1 bits in the signal source are largely random, the resulting RIS time-coded signal exhibits unstable harmonic components in its spectrum. Consequently, directly applying harmonic beamforming methods \cite{8741154, 9573342, 9573342, 10994471, 11053234} designed for periodic modulation is not advisable. Nevertheless, from (\ref{HK}), it can be observed that time coding does not affect the phase of the fundamental component ($k=0$), i.e. the component at frequency $f_c$. When $F_p$ is much smaller than the RF signal frequency $f_c$ incident on the RIS, the space coding corresponding to beamforming at frequency $f_c$ is approximately the same as that at $f_c + F_p$. Based on this observation, this paper proposes a space-time joint coding strategy that simultaneously enables harmonic-based direct modulation communication at  $f_c + F_p$ and beamforming for bit error rate reduction.


\subsection{Space-time Joint Coding Scheme}
This subsection details the operational principle of the proposed space–time joint coding strategy. 

Let the static (initial) state of the $(m,n)$th RIS unit be denoted by
$\Gamma_{m,n}\in\{0,1\}$, which we refer to as the \emph{space coding}. The time-domain modulation applied to each element is denoted by the binary sequence $G(t)\in\{0,1\}$, which we refer to as the \emph{time coding}. The feeding single-tone is
\begin{equation}
	\label{st}
	S(t)=e^{j2\pi f_c t},
\end{equation}
where $f_c$ is the carrier frequency. Ignoring amplitude variations due to path loss (i.e., assuming $|r_{m,n}|\approx\text{const}$ for all elements), the signal received at direction $(\theta,\varphi)$ can be written as
\begin{equation}
	\label{srt}
	\begin{aligned}
		S_r(\theta,\varphi,t)
		&=S(t)\sum_{m=1}^{M}\sum_{n=1}^{N} e^{jK r_{m,n}} e^{j\big[\Gamma_{m,n}\pi + G(t)\pi\big]} w_{m,n},
	\end{aligned}
\end{equation}
where $K=2\pi/\lambda$ is the wavenumber, $\lambda=c/f_c$ is the wavelength, $c$ is the speed of light, and $r_{m,n}$ denotes the propagation distance from the $(m,n)$th element to the receiver. The spatial steering factor $w_{m,n}$ (geometric phase delay relative to the array center) is given by
\begin{equation}
	\label{wmn}
	\begin{aligned}
		w_{m,n}
		&=\exp\Big\{-jK d\Big[\Big(m-\frac{M+1}{2}\Big)\sin\theta\cos\varphi\Big]\Big\} \\
		&\qquad\qquad\times\exp\Big\{-jK d\Big[\Big(n-\frac{N+1}{2}\Big)\sin\theta\sin\varphi\Big]\Big\},
	\end{aligned}
\end{equation}
where $d$ is the element spacing and $j=\sqrt{-1}$.

As shown in Subsection II-A, the zeroth-order Fourier coefficient ($k=0$) of a time-coded sequence carries no phase offset; therefore the carrier-phase response at $f_c$ is not altered by the time coding. In other words, time coding and space coding are (to first order at the carrier) separable: the \emph{time coding} $G(t)$ determines the information-bearing modulation (appearing on the sidebands, e.g., at $f_c + f_p$), while the \emph{space coding} $\Gamma_{m,n}$ determines the static phase distribution used for beamforming at $f_c$. Moreover, when $f_p\ll f_c$ (e.g., $f_p=1\,$MHz and $f_c=3.5\,$GHz), we have $f_c+f_p\approx f_c$, and the spatial phase distribution designed at $f_c$ is approximately valid at the first-order harmonic $f_c+f_p$ for a given beam direction.

In direct modulation operation, the information bits are first mapped to constellation symbols; each transmitted symbol then determines a set of time-coding parameters (for example, duty cycle $\mu_l$ and starting instant $t_{\text{on},l}$), which in turn define the modulation timing sequence $G(t)$. To recover the transmitted symbols at the receiver, a typical procedure is \cite{ni2022reconfigurable,chen2022accurate,yan2024metasurface}:
\begin{enumerate}
	\item extract the instantaneous phase of the received signal;
	\item perform a Fourier transform of the extracted phase time series;
	\item recover the symbols by demodulating the amplitude and phase of the first-order harmonic (e.g., at $f_c+f_p$).
\end{enumerate}

From \eqref{srt} it follows that the information-bearing sideband components are produced solely by the time coding $G(t)$, and are independent of the static space coding $\Gamma_{m,n}$. Hence $\Gamma_{m,n}$ can be exploited to shape the radiated beam (increase SNR at the receiver) without corrupting the direct-modulation information carried by $G(t)$. Based on the above separation property, we implement a joint space–time coding by combining $\Gamma_{m,n}$ and $G(t)$ element-wise using an exclusive-or (XOR) operation. Denote the joint coding applied to the $(m,n)$-th element by $\phi_{m,n}(t)\in\{0,1\}$. Then
\begin{equation}
	\phi_{m,n}(t)=\Gamma_{m,n}\oplus G(t),
	\qquad
	\text{element phase }=\pi\phi_{m,n}(t).
\end{equation}
Table~\ref{Tab-1} summarizes the mapping between space coding, time coding and the resulting element phase.

\begin{table}[ht]
	\caption{Joint space–time coding mapping (XOR).}
	\label{Tab-1}
	\centering
	\begin{tabular}{cccc}
		\hline
		$\Gamma_{m,n}$ & $G(t)$ & $\phi_{m,n}(t)=\Gamma_{m,n}\oplus G(t)$ & Phase (rad) \\ \hline
		0 & 0 & 0 & $0$ \\
		0 & 1 & 1 & $\pi$ \\
		1 & 0 & 1 & $\pi$ \\
		1 & 1 & 0 & $0$ \\ \hline
	\end{tabular}
\end{table}

In order to point the beam of RIS to $(\theta, \varphi)$, the corresponding space coding ${{\Gamma }_{m,n}}$ is calculated by (\ref{gammnmao}) and (\ref{gammn}).

\begin{equation}	
	\label{gammnmao}
	\begin{aligned}
		\tilde{\Gamma }_{m,n}&=Kd (m-\frac{M+1}{2})\sin \theta \cos \varphi \\ &+Kd(n-\frac{N+1}{2})\sin \theta \sin \varphi - kr_{m,n},
	\end{aligned}
\end{equation}
For the 1-bit phase reconfigurable RIS, after performing 1-bit quantization on $\tilde{\Gamma }_{m,n}$, it can be expressed as
\begin{equation}
	\resizebox{0.65\hsize}{!}{$	
		\label{gammn}
		{{\Gamma }_{m,n}}=\left\{ \begin{matrix}
			0,0\le \,\bmod \,\left( {{{\tilde{\Gamma }}}_{m,n}},2\pi  \right)<\pi   \\
			1,\pi \le \,\bmod \,\left( {{{\tilde{\Gamma }}}_{m,n}},2\pi  \right)<2\pi   \\
		\end{matrix} \right.,
		$}
\end{equation}
where mod$(x,y)$ denotes the remainder of $x$ to $y$.

In summary, the proposed XOR-based joint coding enables simultaneous realization of (i) direct modulation communication via the time code $G(t)$ (symbols recovered from sideband harmonics) and (ii) beamforming via the static space code $\Gamma_{m,n}$ (which steers and concentrates energy at the carrier and, under $f_p\ll f_c$, at nearby harmonics). This separation motivates efficient design procedures in which $\Gamma_{m,n}$ is chosen to maximize receive SNR while $G(t)$ is chosen to realize the desired modulation constellation and symbol mapping.


\subsection{Design Procedure}
Based on the separation property discussed above, the design of the joint coding follows a two-step process:  
\begin{enumerate}
	\item \textbf{Space coding:} For a given receiver location $(\theta,\varphi)$, compute the phase distribution using (\ref{gammnmao}) and quantize it to obtain $\Gamma_{m,n}$ as in (\ref{gammn}). This ensures that the RIS beam is steered toward the receiver with maximum array gain.  
	\item \textbf{Time coding:} Map the transmitted bit stream into modulation symbols. Each symbol specifies the duty cycle $\mu_l$ and starting instant $t_{\text{on},l}$ of the on–off sequence $G(t)$, thereby generating the desired communication waveform.  
\end{enumerate}

Finally, the XOR operation is applied element-wise between $\Gamma_{m,n}$ and $G(t)$, yielding the joint coding sequence $\phi_{m,n}(t)$. This process guarantees simultaneous beamforming and symbol modulation without the need for additional RF chains.

\subsection{FPGA-Based Implementation}
The proposed joint coding scheme is realized on a field-programmable gate array (FPGA), which provides real-time control of the RIS elements. The implementation consists of the following modules:
\begin{itemize}
	\item \textbf{Bitstream parser:} Converts the input data into a binary stream and maps it to the corresponding constellation symbols.
	\item \textbf{Time-coding generator:} Produces the sequence $G(t)$ according to the duty cycle and symbol timing parameters associated with each constellation symbol.
	\item \textbf{Space-coding memory:} Stores the precomputed $\Gamma_{m,n}$ patterns for different beam directions. For a given receiver location, the appropriate codeword is retrieved.
	\item \textbf{XOR logic:} Combines $G(t)$ and $\Gamma_{m,n}$ element-wise to generate $\phi_{m,n}(t)$.
	\item \textbf{RIS driver:} Converts $\phi_{m,n}(t)$ into control voltages for the 1-bit phase shifters of the RIS array.
\end{itemize}

The modular architecture enables flexible adaptation to different modulation schemes (e.g., BPSK, QPSK, 16QAM) and beam directions by reprogramming the FPGA. Moreover, the use of pre-stored space-coding patterns ensures fast reconfiguration for dynamic beam steering.

\subsection{System Operation}
During operation, the RIS is illuminated by a continuous-wave carrier $S(t)$. The FPGA synchronously drives each element of the RIS according to $\phi_{m,n}(t)$, such that the joint coding is realized in real time. At the receiver side, the information is recovered by extracting the phase of the received signal, performing spectral analysis, and demodulating the sideband corresponding to $f_c+f_p$, as outlined in the decoding procedure.

\section{Numeric Results}
\begin{figure}[!t]
	\centering
	\subfloat[]{\includegraphics[width=4cm]{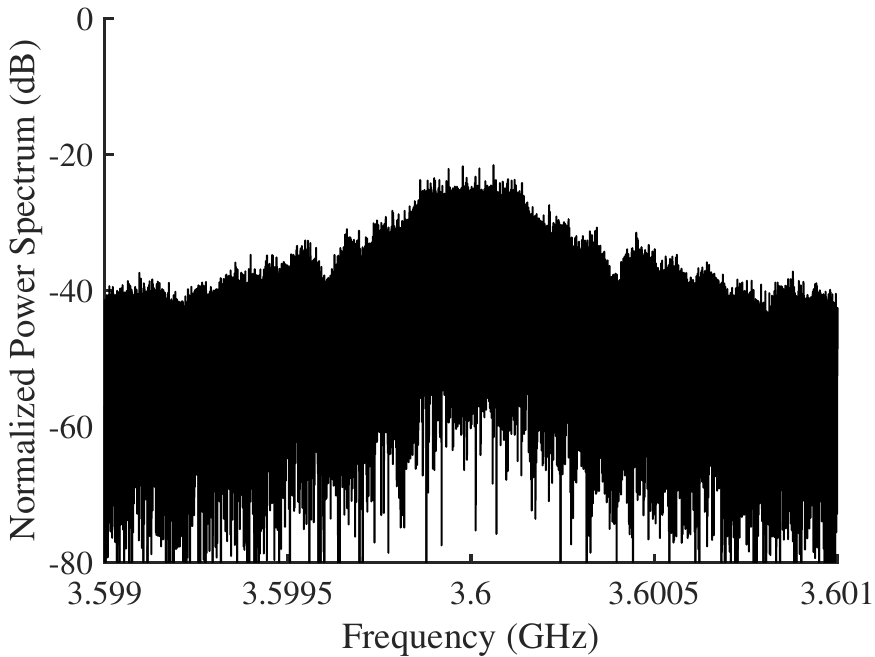}%
		\label{3a}}
	\hfil
	\subfloat[]{\includegraphics[width=4cm]{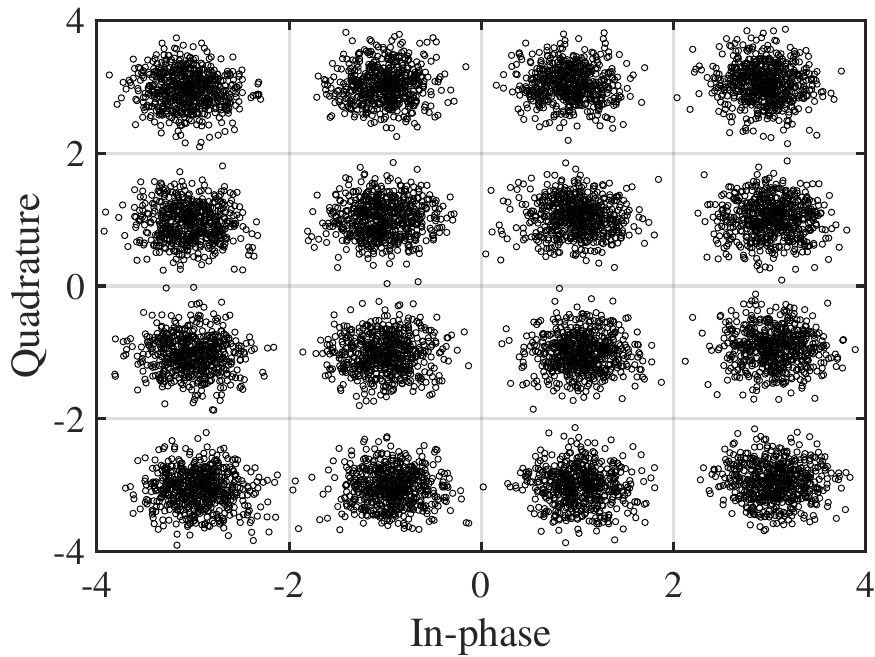}%
		\label{3b}} \\
	\subfloat[]{\includegraphics[width=4cm]{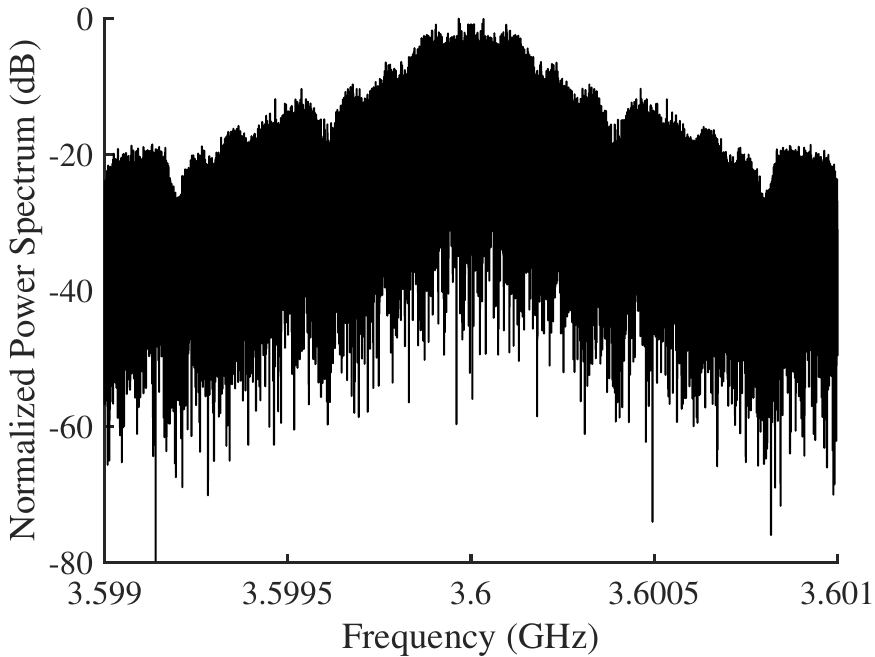}%
		\label{3c}}
	\hfil
	\subfloat[]{\includegraphics[width=4cm]{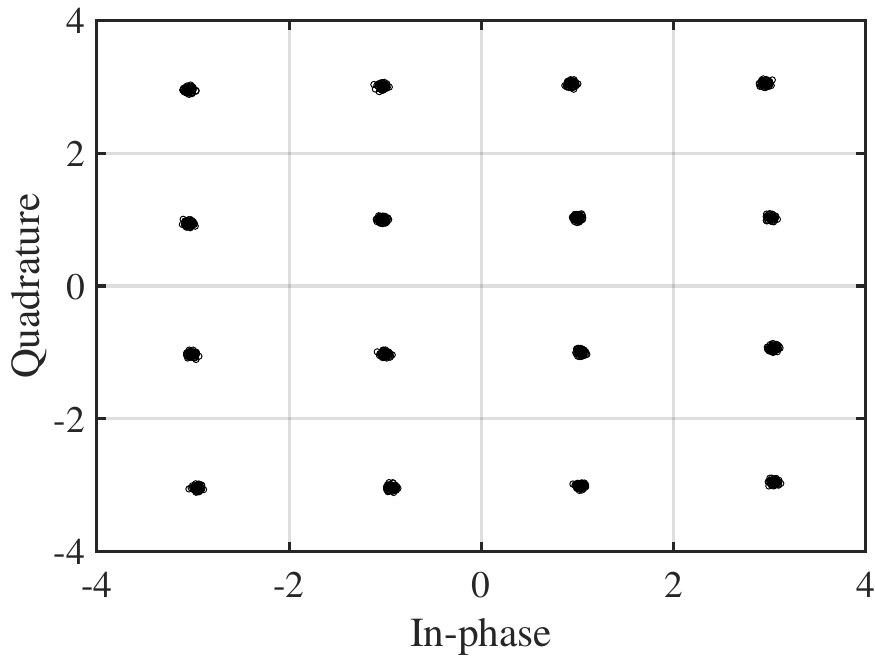}%
		\label{3d}} 	
	\caption{Simulated received signal before and after applying joint space--time coding: (a) spectrum and (b) constellation with time-only coding; (c) spectrum and (d) constellation with joint space--time coding.}
	\label{Fig3}
\end{figure}


This section presents numerical simulations to verify that the proposed joint space--time coding can simultaneously realize beamforming and direct modulation communication. By performing an XOR operation between the space coding and the time coding, the received signal quality is significantly improved due to the enhanced SNR.


Consider a $16\times16$ RIS operating at 3.6~GHz. The simulation parameters are summarized in Table~\ref{Tab-Sim-I}. The RIS switching rate is set to $F_p=100$~kHz and the sampling rate to 20~MSps. The feeding antenna is located at $(0,0,-200)$~mm, while the receiver is positioned at $(\theta,\varphi)=(30^\circ,0^\circ)$ relative to the RIS.
\begin{table}[htbp]
	\caption{Parameters for Numeric Simulation}
	\label{Tab-Sim-I}
	\centering
	\begin{tabularx}{0.5\textwidth}{
			>{\centering\arraybackslash}X 
			>{\centering\arraybackslash}X 
		}
		\toprule
		\textit{Parameters} & \textit{Values} \\
		\midrule
		RIS size & $16\times16$ \\
		Carrier frequency $ F_c $ & 3.6 GHz \\ 
		Modulation frequency $F_p$ & 100 kHz \\
		Sample rate & 20 MSps \\	
		Unit spacing $d$ & 43 mm \\
		$\left( {{x}_{c}},{{y}_{c}},{{z}_{c}} \right)$  & (0,0,-200) mm \\
		$\left( \theta ,\varphi  \right)$ & $(30,0)^\circ$ \\
		SNR  & 10 dB \\
		\bottomrule
	\end{tabularx}
\end{table}
Assume that the direct modulation system adopts 16QAM for symbol mapping. The corresponding time coding sequence $G(t)$ is generated according to the binary source bits.  

\subsection{Received Signal Spectrum and Constellation}

When only time coding is applied (i.e., all RIS elements are switched synchronously), the received signal power spectrum and constellation diagram are shown in Fig.~\ref{Fig3}(a) and Fig.~\ref{Fig3}(b), respectively.  
When the space coding $\Gamma_{m,n}$ is designed according to (\ref{gammnmao}) and (\ref{gammn}) to steer the beam toward the receiver, and the joint space--time coding is generated as in Table~\ref{Tab-1}, the received signal spectrum and constellation are shown in Fig.~\ref{Fig3}(c) and Fig.~\ref{Fig3}(d). 

It is observed that joint space--time coding increases the received power by approximately 22~dB compared with time-only coding. Furthermore, the symbol clusters in the constellation diagram become more distinguishable, indicating improved communication reliability.

\subsection{BER Performance}
\begin{figure}[!t]
	\centering
	\includegraphics[width=8cm]{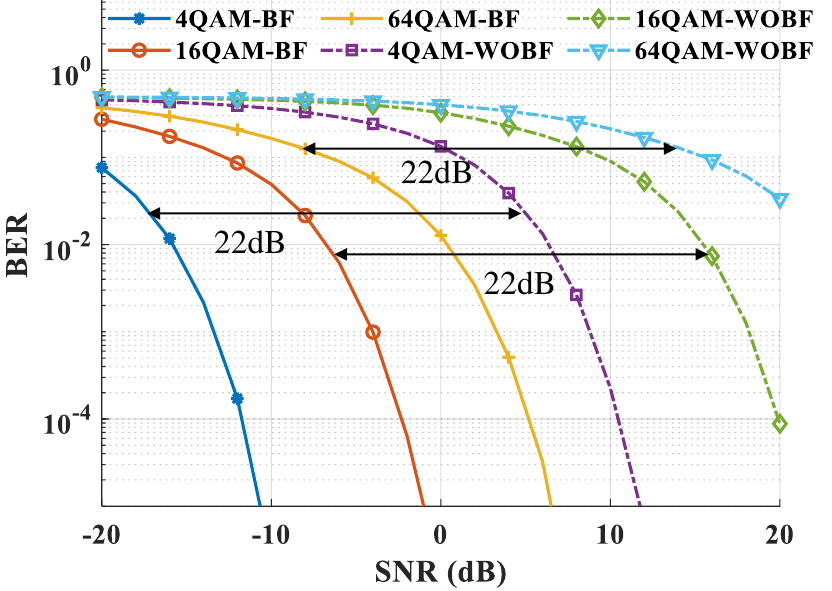}
	\caption{BER versus SNR for different modulation orders with and without beamforming using joint space--time coding.}
	\label{Fig4}
\end{figure}

The BER performance is evaluated under different modulation formats (4QAM, 16QAM, and 64QAM). As shown in Fig.~\ref{Fig4}, the solid lines represent the results with beamforming (BF) using joint coding, while the dashed lines correspond to the case without beamforming (WOBF). It can be seen that, for the same BER, the required SNR is reduced by about 22~dB when joint space--time coding is employed. Additionally, as the modulation order increases, the BER degrades at a fixed SNR, consistent with classical communication theory.

\subsection{Beam Pattern Analysis}

\begin{figure}[t!]
	\centering
	\includegraphics[width=8cm]{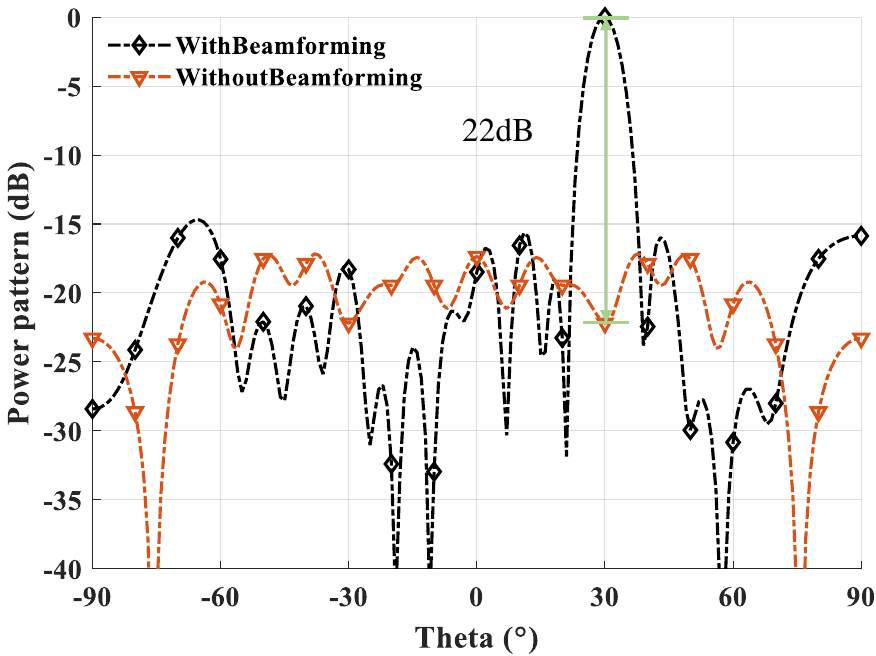}
	\caption{Power pattern of the transmitter before and after beamforming using joint space-time coding.}
	\label{Fig5}
\end{figure}

\begin{figure*}[!t]
	\centering
	\includegraphics[width=12cm]{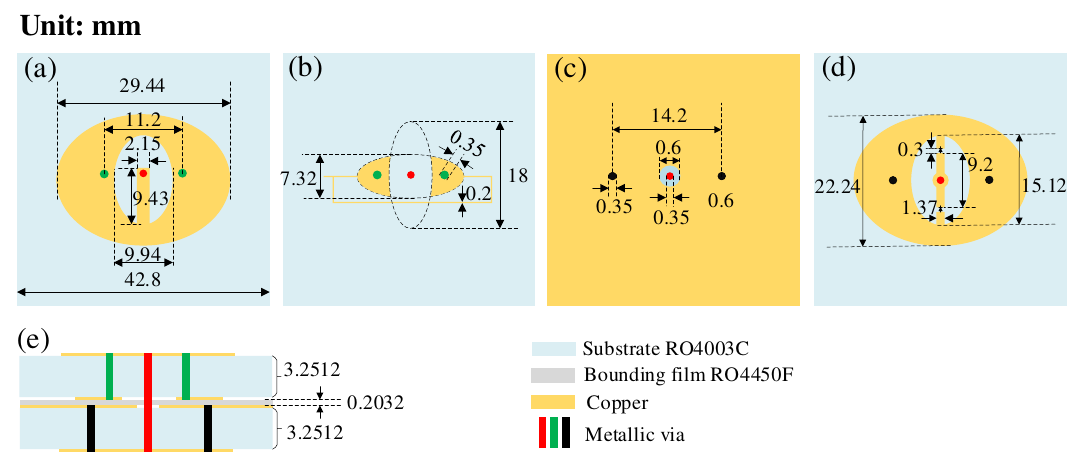}
	\caption{Transmission amplitude and transmission phase of the RIS unit when the unit is in the `ON' state and the `OFF' state, respectively, and the transmission phase difference between these two states.}
	\label{FigUnit}
\end{figure*}



When only time coding is applied, all RIS elements are synchronized, and the instantaneous transmit power pattern can be expressed as
\begin{equation}
	\label{Pthetaphi}
	{P}(\theta ,\varphi)= \sum\limits_{m=1}^{M}{\sum\limits_{n=1}^{N}{\frac{{{e}^{jk{{r}_{m,n}}}}}{\left| {{r}_{m,n}} \right|}{{e}^{j {{\Gamma }_{m,n}}\pi}}{{w}_{m,n}}}},
\end{equation}
The simulated transmit power patterns are plotted in Fig.~\ref{Fig5}. It is evident that beamforming with joint space--time coding enhances the received signal power by approximately 22~dB in the target direction $(30^\circ,0^\circ)$. This explains the improvements in both constellation clarity (Fig.~\ref{Fig3}) and BER performance (Fig.~\ref{Fig4}).

\section{Design of RIS}
To experimentally validate the proposed approach, a 1-bit phase-reconfigurable transmissive RIS operating in the 3.4--3.79~GHz band is designed and fabricated.

The designed transmissive RIS unit mainly consists of four metal layers, two dielectric layers and one adhesive layer, and the layer stack structure is shown in Fig.~\ref{FigUnit}(e). From top to bottom, are the receiving patch, the bias layer, the metal ground and the transmitting layer, as shown from Fig.~\ref{FigUnit}(a) to Fig.~\ref{FigUnit}(d), respectively. The top and bottom layers of substrates consisted of two layers of Rogers RO4003C (dielectric constant of 3.55, loss tangent of 0.0027, single layer with a thickness of 1.524 mm) sandwiched by a bonding film Rogers4450F (dielectric constant of 3.52, loss tangent of 0.004, thickness of 0.2032 mm). Therefore, The overall substrate thickness is therefore 3.2512~mm. As shown in Fig.~\ref{FigUnit}(a), the receiving layer consists of an elliptical metal patch, and a U-slot. As shown in Fig.~\ref{FigUnit}(d), the transmitter layer consists of an elliptical metal patch of the same size as the receiver layer, and two PIN diodes (MADP-000907-14020) with the same orientation. The PIN diode is modeled as a lumped element, which is equivalent to a circuit with a 7 ohm resistor and 0.03 nH inductor in series when it is switched ON and 0.025 pF and 0.03nH in series when it is switched OFF.

\begin{figure}
	\centering
	\includegraphics[width=7cm]{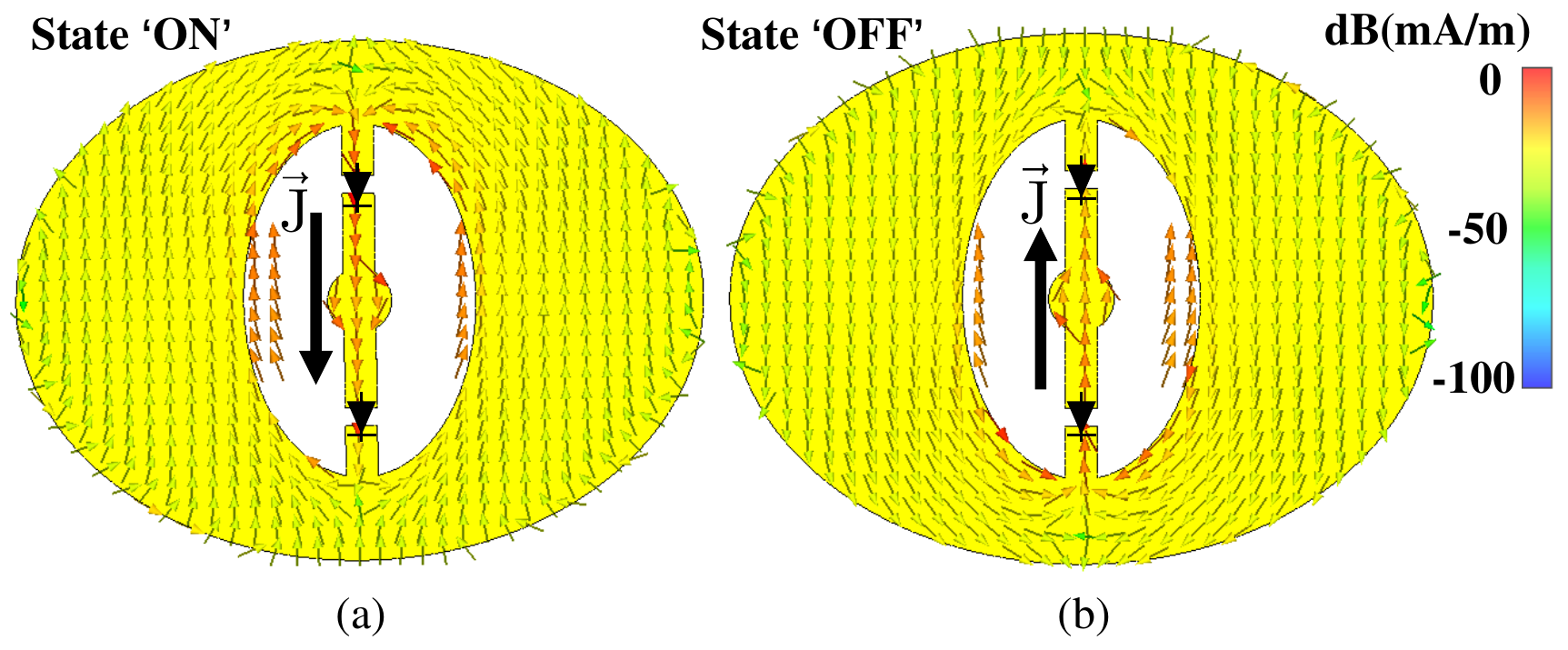}
	\caption{Bottom patch surface current when the unit of RIS is in (a) `ON' state and (b) `OFF' state.}
	\label{FigSurfaceCurrent}
\end{figure}
\begin{figure}
	\centering
	\includegraphics[width=8cm]{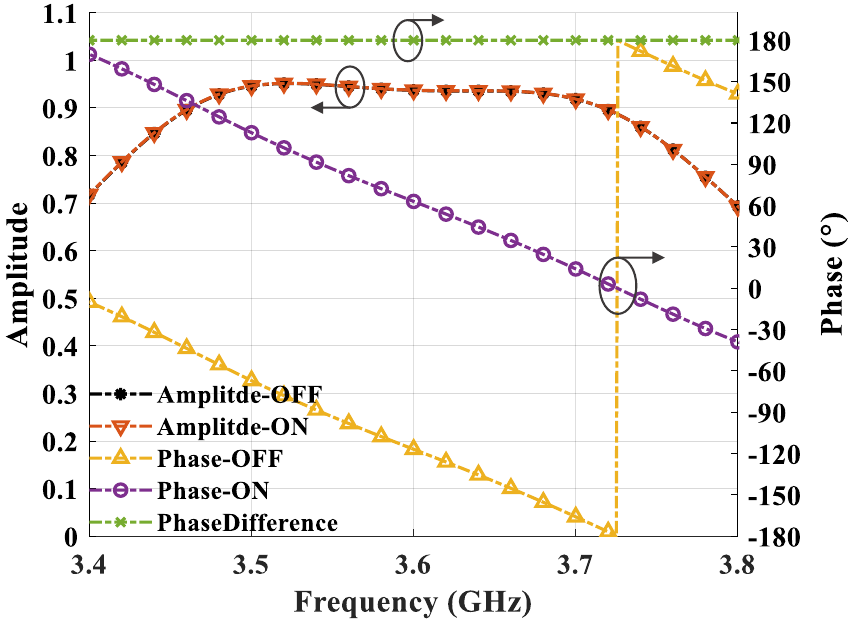}
	\caption{Transmission amplitude and transmission phase of the RIS unit when the unit is in the `ON' state and the `OFF' state, respectively, and the transmission phase difference between these two states.}
	\label{Fig6}
\end{figure}

\begin{figure}
	\centering
	\includegraphics[width=6cm]{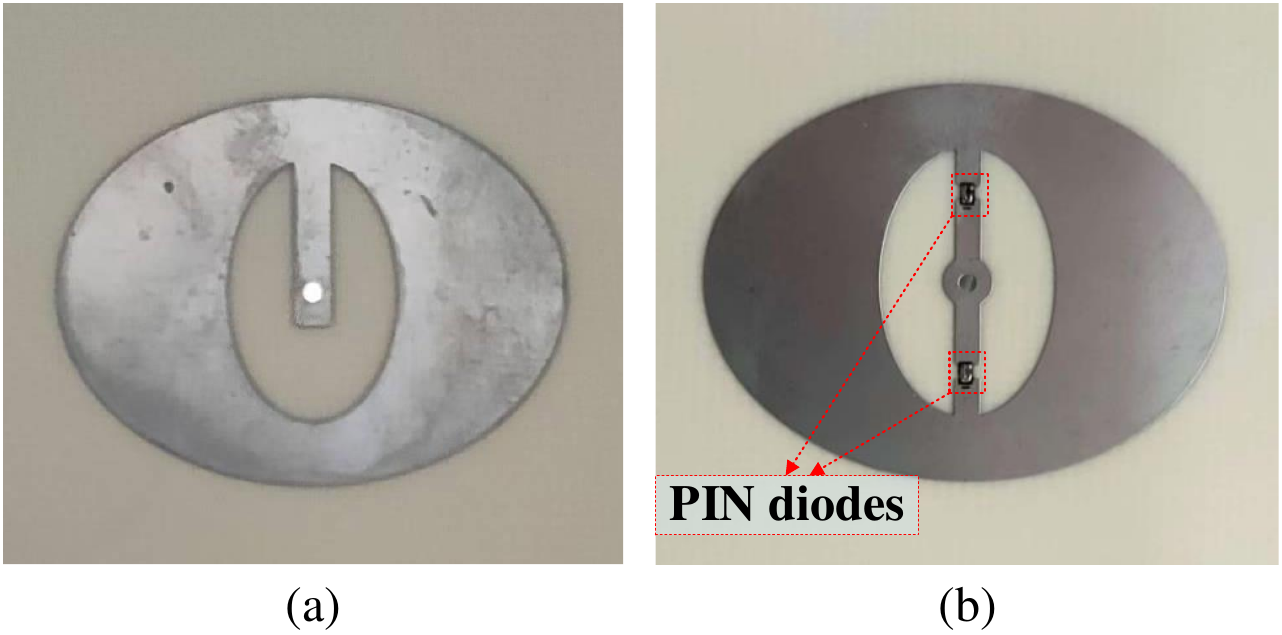}
	\caption{Photos of the RIS unit}
	\label{FigUnitreal} 
\end{figure}

\begin{figure}[t!]
	\centering
	\includegraphics[width=8cm]{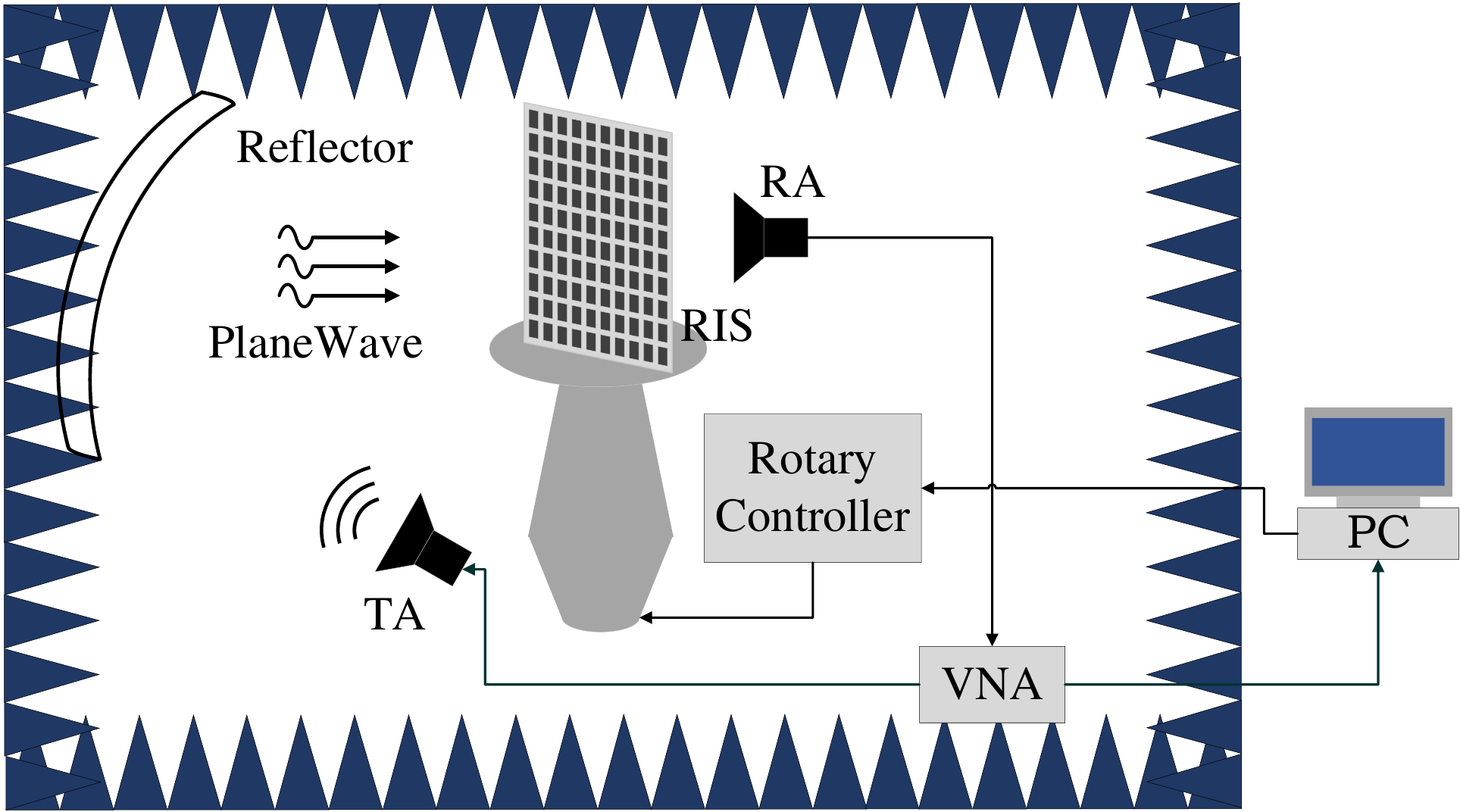}
	\caption{Schematic illustration of the testing scenario for the power pattern on the transmitter end within a direct modulation communication system.}
	\label{Fig7}
\end{figure}

\begin{figure}[t!]
	\centering
	\includegraphics[width=9cm]{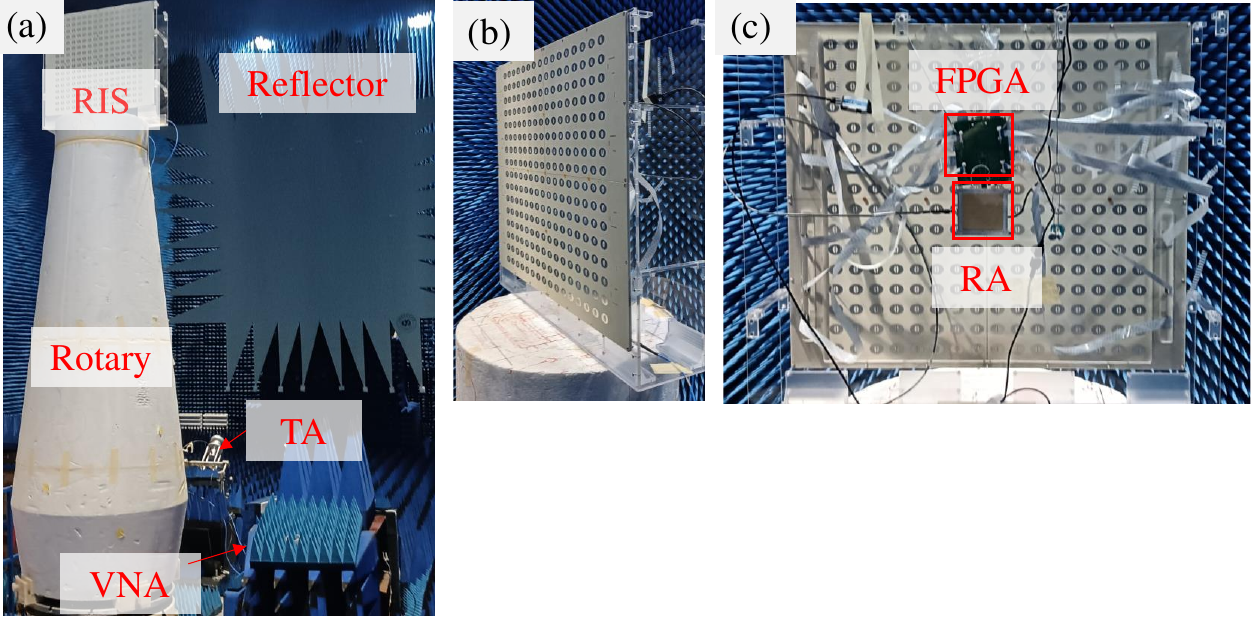}
	\caption{Experimental scenario for testing the power pattern on the transmitter end within a direct modulation communication system.}
	\label{Fig8}
\end{figure}

As illustrated in Fig. 6(a), (b), and (e), the receive patch is connected to the bias layer through metalized vias, and the voltage on the bias line is provided by FPGA. As shown in Fig. 6(c), (d) and (e), the transmit layer is connected to the ground through a metalized vias. The metallized vias connected between the transmit patch and the receive patch are used for RF energy transmission. Subsequently, in the scenario where the voltage on the bias line attains a high level (greater than 1.5 V), the PIN diode positioned at the bottom of the transmitting patch becomes conductive (switches on), while the PIN diode at the top enters a non-conductive state (switches off), thereby placing the unit in the `ON' state. Conversely, when the voltage on the bias line is not at a high level (less than 0.7 V), the PIN diode at the bottom switches off and the PIN diode at the top switches on, consequently setting the unit in the `OFF' state. The switching between the two states causes an inversion of the current on the surface of the transmitter patch \cite{TimeModulated}, which causes a $180^\circ$ shift in the transmission phase, but maintains the same transmission amplitude.  

Full-wave electromagnetic simulation of the designed structure was carried out by using CST Microwave Studio (MWS). When the RIS unit is in the `ON' state and `OFF' state, respectively, the current distribution on the surface of the transmitter patch is shown in Fig.~\ref{FigSurfaceCurrent}, which shows that the current is inverted,
and the corresponding amplitude and phase responses are shown in Fig.~\ref{Fig6}. The difference in phase response between these two states is approximately $180^\circ$, and the transmission amplitude is greater than 0.7, which fulfills the communication requirements in the 3.4-3.79 GHz band.  The receiving patch and transmitting patch of the fabricated RIS unit are presented in Fig.~\ref{FigUnitreal}(a) and Fig.~\ref{FigUnitreal}(b), respectively.

\begin{figure}[t!]
	\centering
	\includegraphics[width=6cm]{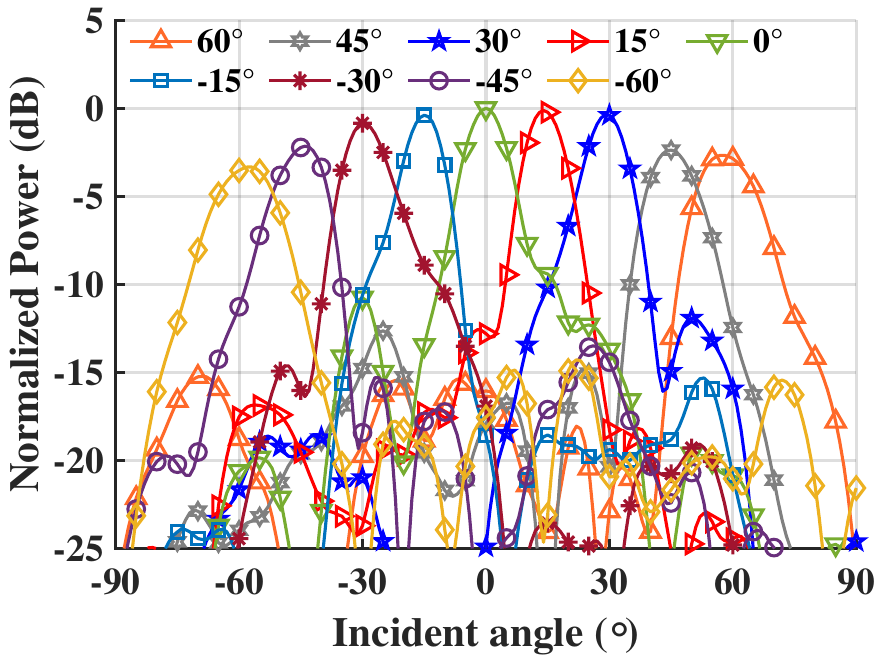}
	\caption{Beam scanning results at the transmitter of a directly modulated communication at 3.6 GHz.}
	\label{FigBeamScaning}
\end{figure}

\begin{figure}[t!]
	\centering
	\includegraphics[width=6cm]{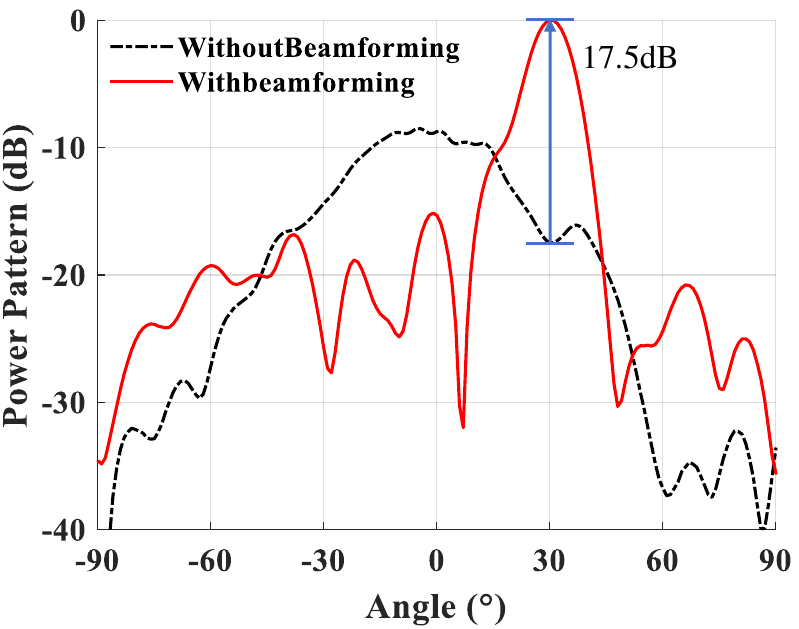}
	\caption{The measured power patterns when all units of the RIS are set to be in the same state (simulating the use of time coding only without space coding) and all units are space coded in accordance with the 30° beam pointing (simulating the use of space-time joint coding), respectively.}
	\label{Fig9}
\end{figure}

\section{Experimental Results}

\begin{figure*}[t!]
	\centering
	\includegraphics[width=10cm]{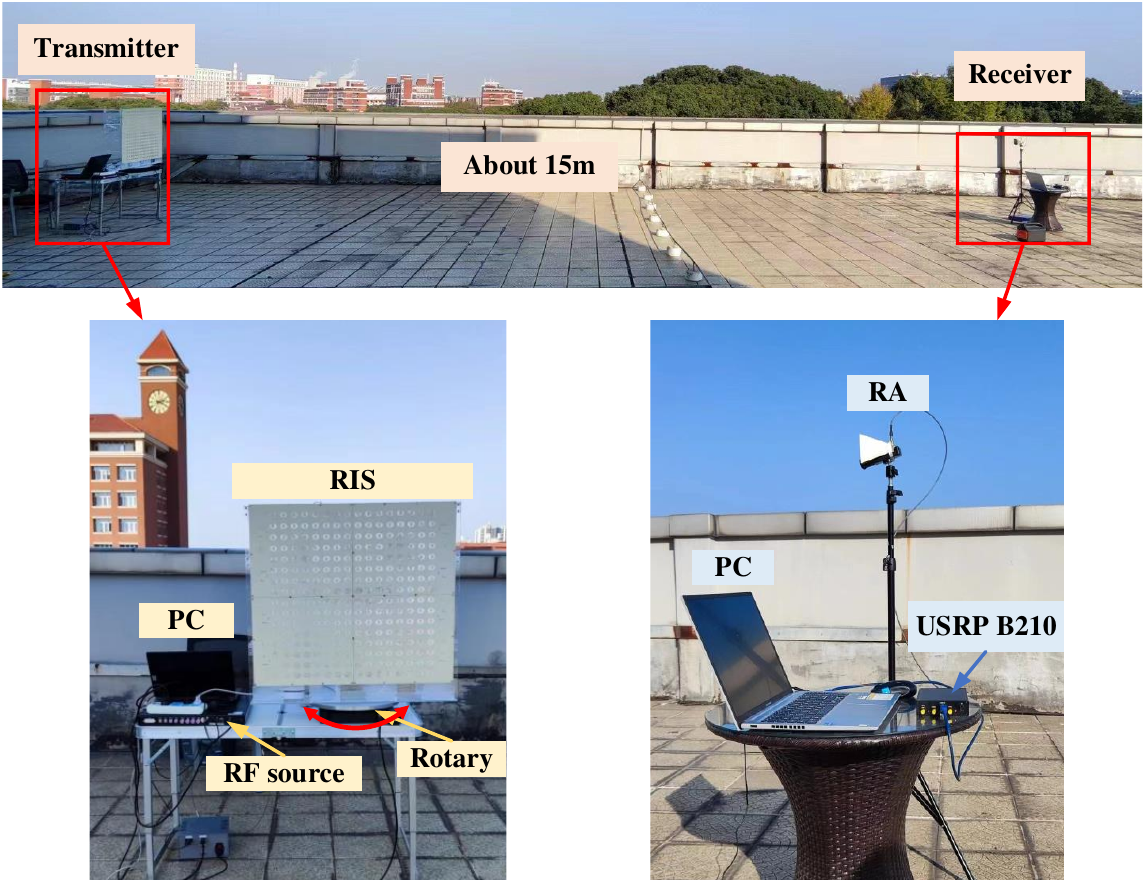}
	\caption{Experimental scenarios for direct modulation communication.}
	\label{Fig10}
\end{figure*}

This section presents the experimental validation of the proposed method for enhancing direct modulation communication through joint space-time coding on reconfigurable intelligent surfaces (RIS). 

\subsection{Measurement of Beamforming}

The first experiment focuses on evaluating the beamforming performance of the RIS in the direct modulation communication system. Specifically, we aim to assess the power pattern at the transmitter end under different configurations of space-time coding. The experimental setup for this measurement is depicted in Fig.~\ref{Fig7}. The RIS and its feeding antenna (receive antenna) are placed in the quiet zone of a microwave anechoic chamber. To characterize the power pattern, the S21 parameters between the transmitting antenna (TA) and the receiving antenna (RA) are measured at various angles using a vector network analyzer (VNA). This allows us to observe the distribution of the transmitted signal power at the receiver.

The beam scanning capability of the RIS is first tested at 3.6 GHz. For this purpose, space coding corresponding to different beam steering angles is applied to the RIS, and the power pattern is measured for each configuration. The experimental results, shown in Fig.~\ref{Fig9}, reveal that the peak angle of the measured power pattern closely matches the expected beam direction. This confirms that the space coding applied to the RIS enables successful beam steering, which is essential for the proposed joint space-time coding method.

To further evaluate the impact of space-time joint coding on beamforming, we compare the power patterns under two different configurations: (1) when all RIS units are set to the same state, simulating the case where only time coding is used (no space coding), and (2) when all RIS units are space coded according to the 30° beam steering direction, simulating the use of space-time joint coding. The measured power patterns for both scenarios are presented in Fig.~\ref{Fig9}. It can be seen that after applying space-time joint coding, the receiver located at the 30° direction experiences a significant power gain of approximately 17.5 dB. This result highlights the advantage of combining beamforming and direct modulation communication via joint space-time coding. 

For the sake of clarity, we note that the measured gain is slightly lower than the theoretical 22 dB mentioned in Fig.~\ref{Fig5}. This discrepancy is attributed to the fact that the RIS and feeding antenna are modeled as omnidirectional in (\ref{Pthetaphi}), which simplifies the analysis but results in a slightly lower gain in the experimental setup.


\subsection{Measurement of Directional Communication Performance}

\begin{figure}[t!]
	\centering
	\includegraphics[width=6cm]{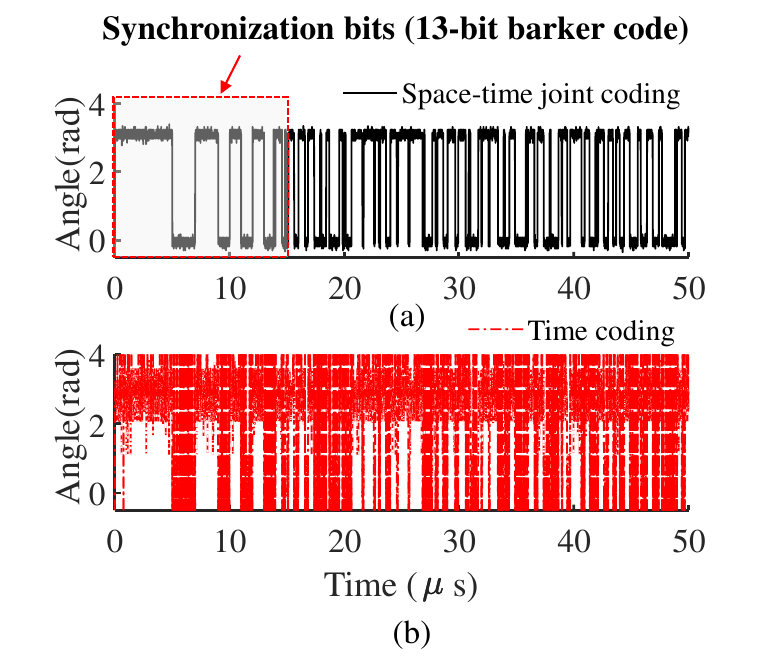}
	\caption{The phase of the received signal (a) under the condition of using only time coding as well as (b) when employing joint time-space coding.}
	\label{Fig11}
\end{figure}

\begin{figure}[!t]
	\centering
	\subfloat[]{\includegraphics[width=4cm]{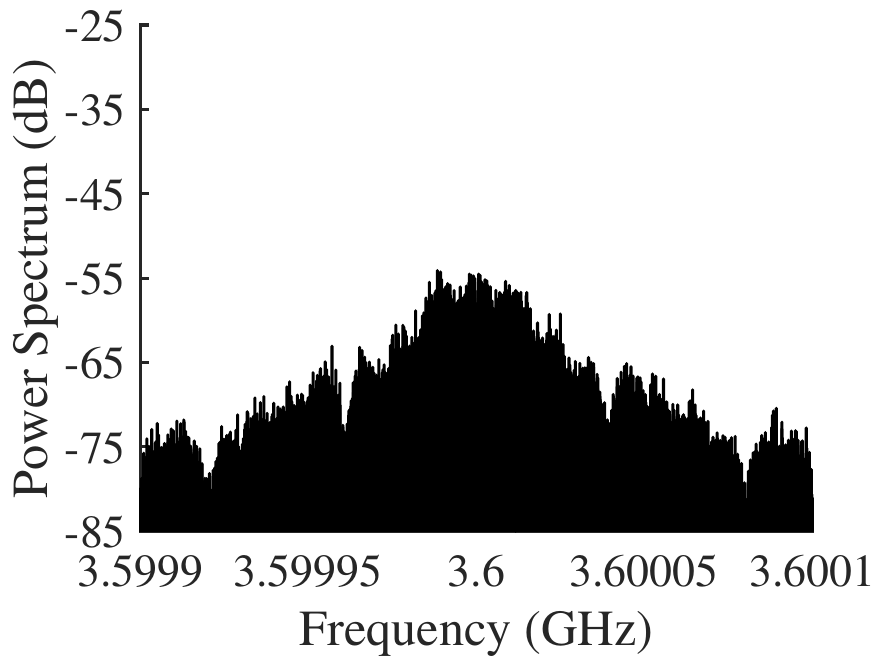}%
		\label{12a}}
	\subfloat[]{\includegraphics[width=4cm]{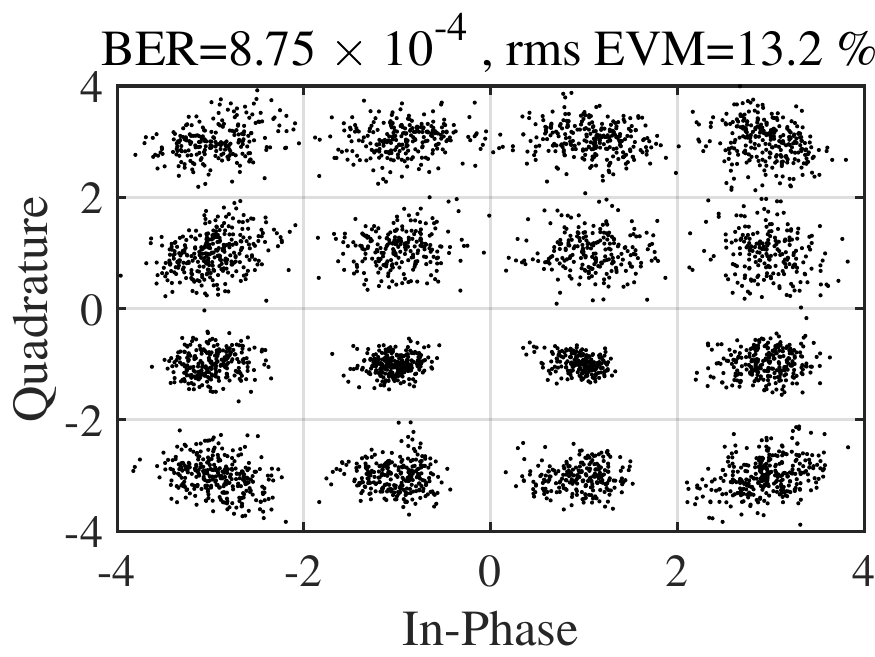}%
		\label{12b}} \\
	\subfloat[]{\includegraphics[width=4cm]{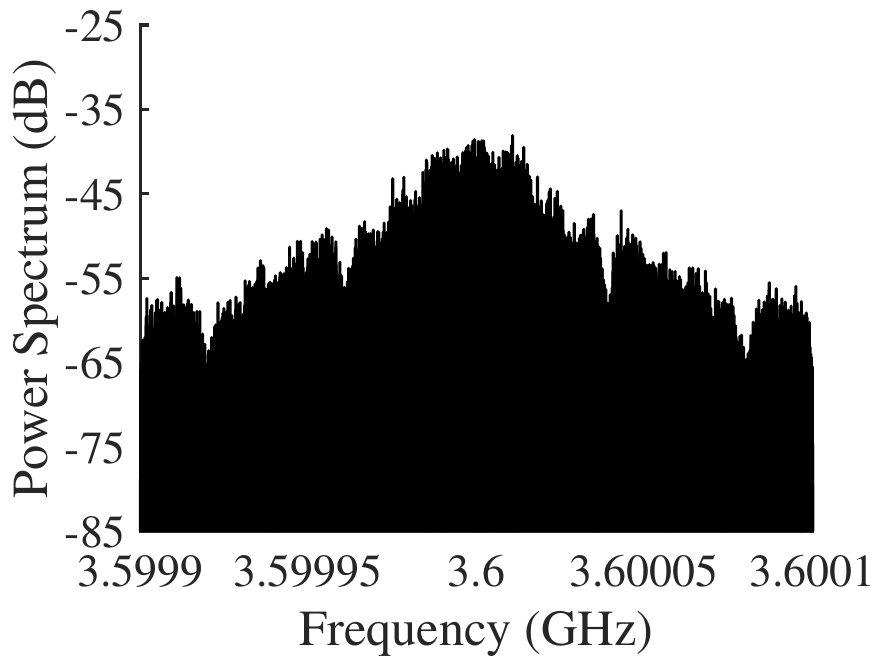}%
		\label{12c}}
	\subfloat[]{\includegraphics[width=4cm]{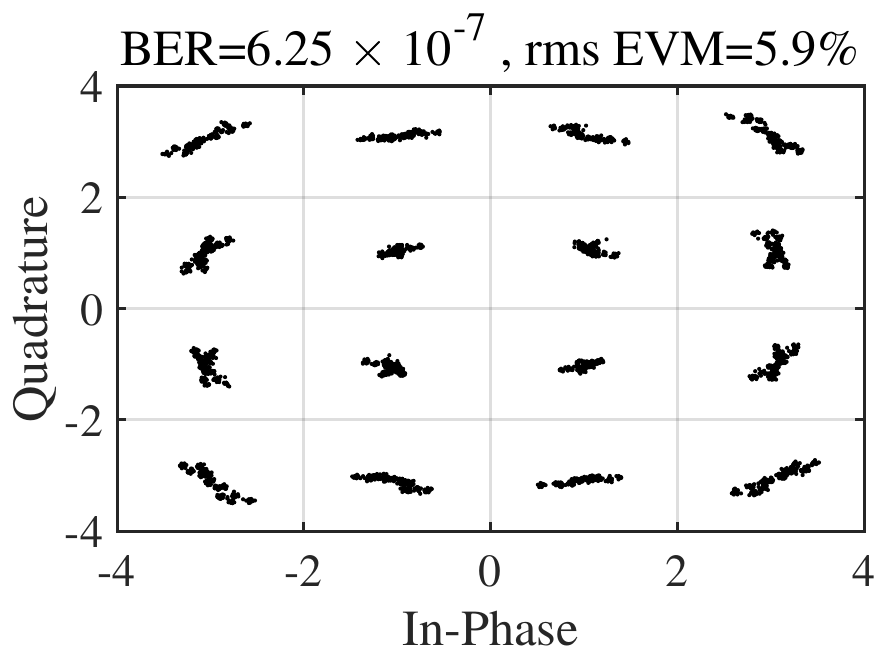}%
		\label{12d}} 	
	\caption{(a) Power spectrum and (b) constellation diagram of the received signal when only time coding is used. (c) Power spectrum and (d) constellation diagram of the received signal when joint time-space coding is applied.}
	\label{Fig12}
\end{figure}

The next experiment evaluates the performance of the direct modulation communication system, focusing on the gain achieved through the use of space-time joint coding. The experimental setup for this scenario is shown in Fig.~\ref{Fig10}, where the transmissive RIS prototype is mounted on a controlled turntable to adjust the declination angle between the transmitter and receiver. The receiving antenna is placed approximately 15 meters away from the center of the RIS. The modulation method used is direct 16QAM, and the source power at the transmitter end is set to 0 dBm. The duration for each RIS state is 1 $\mu$s, corresponding to a bit rate of 1 Mbps.

To simulate realistic transmission conditions, the data is transmitted in frames. Each frame consists of a synchronization code and a data code. For synchronization, a 13-bit Barker code is used, and the phase of the received signal is cross-correlated with the Barker code to determine the starting point of the data frame. It is assumed that the transmission channel remains stable during the transmission of a single frame. This assumption allows us to directly utilize the received signal for synchronization and compensates for any channel-induced amplitude and phase shifts.

In the receiver, the signal acquisition process includes down-conversion and sampling, which is performed using software radio equipment (USRP B210). Symbol demodulation involves intercepting the received data, performing DFT (Discrete Fourier Transform) calculations, and making symbol decisions using a MATLAB program. Additionally, carrier synchronization calibration is conducted in advance to compensate for any carrier offset between the transceiver devices.

The comparative experiments are performed by rotating the turntable so that the receiver is positioned at $(0^\circ, 30^\circ)$ relative to the RIS. Two sets of bit streams are transmitted: one using only time coding and the other using space-time joint coding. The phase of the received signal under both conditions is shown in Fig.~\ref{Fig11}. In Fig.~\ref{Fig11}(a), the phase of the received signal when only time coding is applied is presented. In Fig.~\ref{Fig11}(b), the phase of the received signal with joint time-space coding is shown. From these time-domain waveforms, it is clear that the signal phase with space-time joint coding is much smoother, with fewer fluctuations and glitches compared to the case where only time coding is used. This indicates that space-time joint coding effectively mitigates phase instability and improves the overall signal quality.

In the frequency domain, the power spectrum of the received signal is compared for both cases. As shown in Figs.~\ref{Fig12}(a) and (c), the power of the received signal is approximately 17.5 dB higher when space-time joint coding is employed, compared to the use of time coding alone. This result is consistent with the beamforming power gain observed in Fig.~\ref{Fig9}. Additionally, the constellation diagrams of the received signal are shown in Figs.~\ref{Fig12}(b) and (d). With space-time joint coding, the constellation diagrams are more tightly clustered, with increased symbol spacing. This indicates that the symbols are better separated, leading to improved detection accuracy and a reduction in the likelihood of symbol errors.

To quantify the improvement in communication performance, we calculate the rmsEVM (root-mean-square error vector magnitude) and the BER (bit error rate) for both cases. The results are summarized in Table~\ref{Tab2}. Using space-time joint coding, the rmsEVM is reduced from 13.2\% to 5.9\%, which corresponds to a 55\% improvement. More significantly, the BER is reduced by three orders of magnitude, from $8.75 \times 10^{-4}$ to $6.25 \times 10^{-7}$. This substantial reduction in BER demonstrates the effectiveness of space-time joint coding in improving the reliability of direct modulation communication systems.

These experimental results validate the proposed joint space-time coding technique, demonstrating its ability to enhance the performance of direct modulation communication by improving both signal quality and communication reliability. The measured improvements in SNR, EVM, and BER highlight the practical advantages of integrating beamforming with direct modulation communication using RIS.

\begin{table}[htbp]
	\caption{The rmsEVM and BER of directly modulated communication system.}
	\label{Tab2}
	\centering
	\begin{tabularx}{0.5\textwidth}{
			>{\centering\arraybackslash}X 
			>{\centering\arraybackslash}X 
			>{\centering\arraybackslash}X 
		}
		\toprule
		\text{Modulation of RIS} & \text{rmsEVM (\%)} & \text{BER} \\
		\midrule
		Space-time joint coding & \textbf{5.9 $\downarrow$} & $\mathbf{6.25\times10^{-7}} \downarrow$   \\
		Only time coding  & 13.2 & $8.75\times10^{-4}$ \\ 
		\bottomrule
	\end{tabularx}
\end{table}

These results clearly demonstrate that the proposed joint space-time coding approach significantly improves the performance of RIS-based direct modulation communication systems. The enhancements in SNR, EVM, and BER provide strong evidence for the practical feasibility of this method in real-world communication systems.

\section{Conclusion}
In this paper, we proposed a joint space-time coding scheme for RIS that enables simultaneous beamforming and direct modulation. By performing XOR operations on time coding (for direct modulation) and space coding (for beamforming), a unified space-time coding is created to achieve both functions simultaneously. A transmissive 1-bit phase reconfigurable RIS operating in the S-band was fabricated to implement this method in a direct modulation communication system. Experimental results in an anechoic chamber confirmed that the RIS, along with the feeding antenna, is capable of beam scanning through space coding. Real-world testing further showed that the joint space-time coding not only enables direct modulation communication at rates up to 10 Mbps, but also enhances the SNR at the receiver, reducing both BER and rmsEVM. This approach shows promising potential for IoT and 6G applications. Future work will explore its application in multiuser systems and further rate enhancement.


\bibliographystyle{IEEETran}
\bibliography{Reference}
\end{document}